\documentclass[prl,superscriptaddress,twocolumn]{revtex4}

\usepackage{amsmath}
\usepackage{graphicx}

\newcommand{\phr}{Phys.\ Rep.\ }

\newcommand{\njp}{New Jour.\ Phys.\ }

\newcommand{\job}{Jour.\ Opt.\ B }

\begin{document}

\title{Dynamical enhancement of spatial entanglement in massive particles}

\author{Michael Lubasch}
\affiliation{Max-Planck-Institut f\"ur Quantenoptik, Hans-Kopfermann-Stra\ss{}e 1, 85748 Garching, Germany}
\author{Florian Mintert}
\affiliation{Albert-Ludwigs-Universit\"at Freiburg, Freiburg Institute for Advanced Studies, Albertstra\ss e 19, Physikalisches Institut, Hermann-Herder-Stra\ss{}e 3, 79104 Freiburg, Germany}
\author{Sandro Wimberger}
\affiliation{Institut f\"ur Theoretische Physik, Universit\"at Heidelberg, Philosophenweg 19, 69120 Heidelberg, Germany}

\date{\today}

\begin{abstract}
We discuss dynamical enhancement of entanglement in a driven Bose-Hubbard model and find an enhancement of two orders of magnitude which is robust against fluctuations in experimental parameters.
\end{abstract}

\maketitle

Quantum coherence is often thought to be found only in very small systems or under artificial lab-conditions, since otherwise unavoidable environment coupling results in rapid loss of coherence.
Whereas this holds quite generally, it holds in particular for many-body coherence, synonymous for entanglement.
As recent experimental \cite{exp} and theoretical \cite{theo} evidence suggests, however, exceptions to this general rule exist.
In particular coherent driving can compensate for the environment-induced loss of coherence
and, thereby stabilize entanglement under conditions under which a static system would be completely separable \cite{driven, drivenions}.

Such dynamically induced entanglement does not only hold the potential to influence macroscopically observable properties \cite{ghosh}, but certainly also opens up new paths towards scalable quantum information processing which otherwise is limited through the unfavourably scaling dephasing times with the system size \cite{nielsen, haeffner}.
However, our current understanding of dynamical enhancement of entanglement is still in its infancy.

\begin{figure}[h]
\centering
\includegraphics[width=0.45\textwidth]{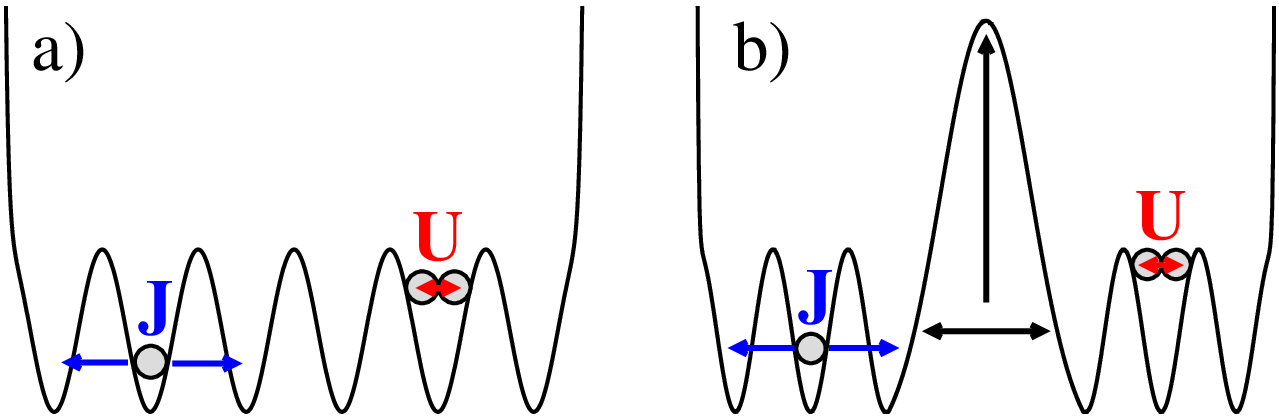}
\caption{\label{fig:model}(color online) a) The system is prepared in an en\-tangled state. b) The system is split into two halves by raising the intermediate barrier. $J$ is the kinetic term and $U$ the on-site interaction in the Bose-Hubbard model \eqref{eq:hwbhm}.}
\end{figure}

In this Letter, we investigate the dynamical enhancement of entanglement between ultracold bosonic atoms stored in an optical lattice that gives rise to a spatially periodic potential created by two counterpropagating laser beams of wavelength $\lambda$ and amplitute $V_0$ in one direction.
A tight perpendicular confinement of strength $V_{\perp}$ in the other two directions restricts the motion of the atoms to one dimension.
In the deep lattice limit \mbox{$V_{0} \gg E_{R}$}, where \mbox{$E_{R} = \hbar^{2} k^{2} / 2 m$} (with \mbox{$k = 2 \pi / \lambda$}) is the recoil energy,
and at sufficiently low temperatures, this system can be well described \cite{bhm} in terms of the Bose-Hubbard Hamiltonian
\begin{eqnarray}\label{eq:hwbhm}
 \hat{H} & = & - J \sum_{l=1}^{L-1} (\hat{a}_{l}^{\dag} \hat{a}_{l+1} +
                                     \hat{a}_{l+1}^{\dag} \hat{a}_{l}) +
                 \frac{U}{2} \sum_{l=1}^{L} \hat{n}_{l} (\hat{n}_{l} - 1)
\end{eqnarray}
where the creation operator $\hat{a}_{l}^{\dag}$ creates and the annihilation operator $\hat{a}_{l}$ annihilates a boson at lattice site $l$.
The tunneling parameter $J$ and the on-site interaction $U$ depend on the lattice parameters approximately \cite{zwerger} via
\begin{eqnarray}
 J / E_{R} & = & \frac{4}{\sqrt{\pi}}( V_{0} / E_{R} )^{\frac{3}{4}} e^{-2 \sqrt{ V_{0} / E_{R}}}\ ,
          \mbox{ and} \\
 U / E_{R} & = & \sqrt{\frac{8}{\pi}} k a_{s} ( V_{0} V_{\perp}^{2} / E_{R}^{3} )^{\frac{1}{4}} \qquad .
\end{eqnarray}
Whereas the lattice depth $V_0$ is typically time-independent, we will compare here the dynamics of such an autonomous system with its driven version \cite{stoeferle},
where $V_0$ is modulated temporally
\begin{eqnarray}\label{eq:V0}
 V_{0}(t) & = & V \Big( 1 + \mathrm{d} V \sin(\omega t) \Big)\ .
\end{eqnarray}
As we will show, the temporal modulation of the tunneling parameter $J$ and the on-site interaction $U$ that results from this lattice depth modulation drives the atoms into a spatially strongly correlated, {\it i.e.}\ entangled state.
For the verification of the entanglement properties, we envision a rapid separation of the few-body system into two parts, what can be realized by ramping up a potential barrier as depicted in Fig.~\ref{fig:model}.

This spatial separation effectively switches off the interaction between the two subsystems what freezes the entanglement dynamics.
However, it also entails that each subsystem will typically {\em not} have a well-defined par\-ticle number.
Whereas correlations in the particle number formally may imply entanglement, its experimental verification will be technically impossible because this would require the measurement of coherent super\-positions of states with different numbers of massive particles \cite{wiseman}.
Therefore, we will postselect \cite{pdc} states with well-defined local particle number.
We will focus in particular on those cases in which particles are split evenly between the two subsystems, since this is the case that occurs with highest probability, and this is also the case in which the highest entanglement can be achieved.
Doing so, we obtain a clean notion of entanglement between the two separated halves of the optical lattice where each half is filled with a fixed number of particles and can be addressed individually.

In the following we quantify the entanglement of the postselected states with the entropy of entanglement \cite{nielsen, bennett} in the case of pure states and the negativity \cite{vidal} in the case of mixed states.
The entropy of entanglement is given by the von Neumann entropy of the reduced density matrix $\varrho_r$ that is obtained through the partial trace over one subsystem of the entire many-body state, {\it i.e.}\ \mbox{$E(\Psi) = -\mbox{Tr} \varrho_r \log_{2} \varrho_r$}.
The negativity \mbox{$N(\rho) = (||\rho^{PT}|| - 1) / 2$} of a mixed state $\varrho$ is defined in terms of the trace norm of the partially transposited density matrix $\rho^{PT}$.

\begin{figure}[h]
\centering
\includegraphics[width=0.45\textwidth]{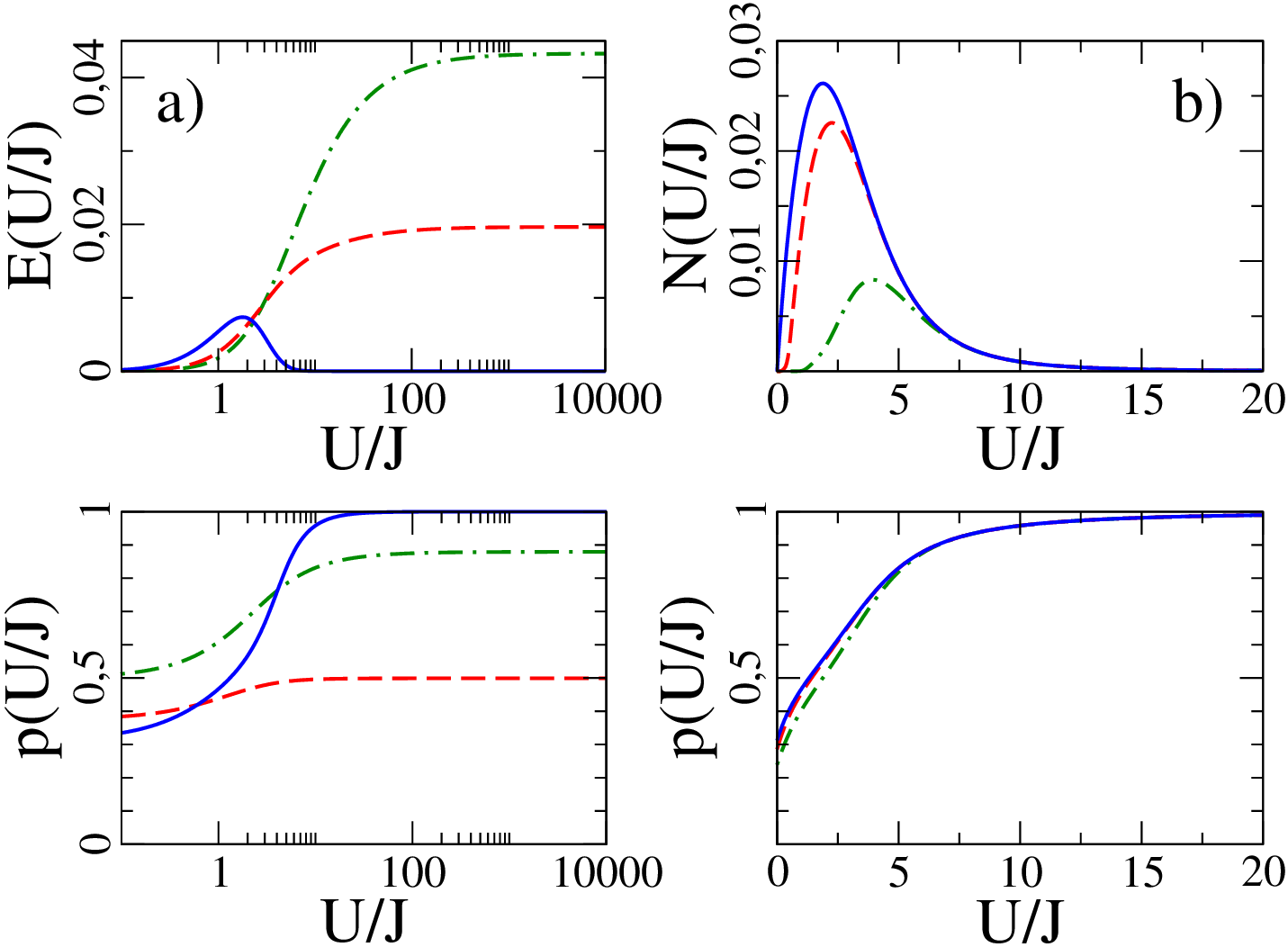}
\caption{\label{fig:gsthpe} (color online) a) Entropy of entanglement $E$ and probability for successful postselection of evenly distributed number of bosons between the right and the left part of a \mbox{$L = 6$} well Bose-Hubbard system.
The overall particle number is \mbox{$N = 6$} (solid blue), \mbox{$N = 3$} (dashed red) and \mbox{$N = 2$} (dash-dotted green).
b) Negativity for the thermal state at temperature \mbox{$T = 0 \,\mathrm{nK}$} (solid blue), \mbox{$T = 40 \,\mathrm{nK}$} (dashed red) and \mbox{$T = 80 \,\mathrm{nK}$} (dash-dotted green) for \mbox{$N = 6$} particles.}
\end{figure}

Fig.~\ref{fig:gsthpe}a) depicts the ground state entanglement proper\-ties for the exemplary case of \mbox{$L = 6$} wells filled with \mbox{$N = 6$}, $3$ and $2$ bosons, respectively.
In all these cases the system features extremely low entanglement, which is also observed for different system sizes and particle numbers.
The qualitative dependence of $E$ on the parameter $U/J$ is as expected:
If the tunneling dominates the system dynamics, {\it i.e.}\ if \mbox{$U/J \simeq 0$}, the bosons populate the same single-particle states such that after postselection of local particle number the system is separable.
For finite interaction $U$ the bosons repel each other and establish correlations;
therefore, $E$ typically increases with $U/J$.
There are, however, exceptions, like the case of unit filling (depicted in blue),
where a separable, perfect Mott insulator \cite{bhm, zwerger} develops for \mbox{$U/J \to \infty$}.
This is also reflected in the fact that in this limit the bosons will always be separated in a balanced fashion between the left and right half of the system,
whereas typically the probability for this is smaller than unity.

Fig.~\ref{fig:gsthpe} shows data for rather small fillings, but also the entanglement properties of the ground state for larger filling \mbox{$N/L > 1$} can be inferred from this data since systems with \mbox{$N + mL$} bosons (with integer $m$) behave qualitatively similar as the system with $N$ bosons.
That is, the maximal amount of entanglement in the ground state does typically not exceed the value of \mbox{$E \approx 0.05$}.

Assuming perfect ground state cooling is certainly a theoretical idealization,
but also thermal excitation cannot enhance the entanglement as shown in Fig.~\ref{fig:gsthpe}b), where the negativity of the thermal state \mbox{$\rho_{th} = \exp(\frac{- \hat{H}}{k_{B} T}) / Z$} is shown.
The probability to find the bosons split evenly into left and right half decreases with increasing temperature,
and the entanglement is even lower than for \mbox{$T = 0$}.

As we will see in the following, this is strongly contrasted by the behavior of the driven system.
To be specific, we consider the experimental parameters: \mbox{$V = 10 \,\mathrm{E_{R}}$} and \mbox{$V_{\perp} = 30 \,\mathrm{E_{R}}$} as lattice depths, \mbox{$\mathrm{d}V = 0.2$} as lattice depth modulation, \mbox{$\lambda = 842 \,\mathrm{nm}$} as the wave length of the laser, \mbox{$a_{s} = 5.45 \,\mathrm{nm}$} as the scattering length and \mbox{$m = 86.909 \,u$} as the mass of rubidium-87 \cite{kollatharimondo}.
As driving frequency, we chose \mbox{$\omega = U / \hbar = 12862 \,\mathrm{Hz}$},
what corresponds to resonant driving in the Mott-insulating regime.

\begin{figure}[h]
\centering
\includegraphics[width=0.45\textwidth]{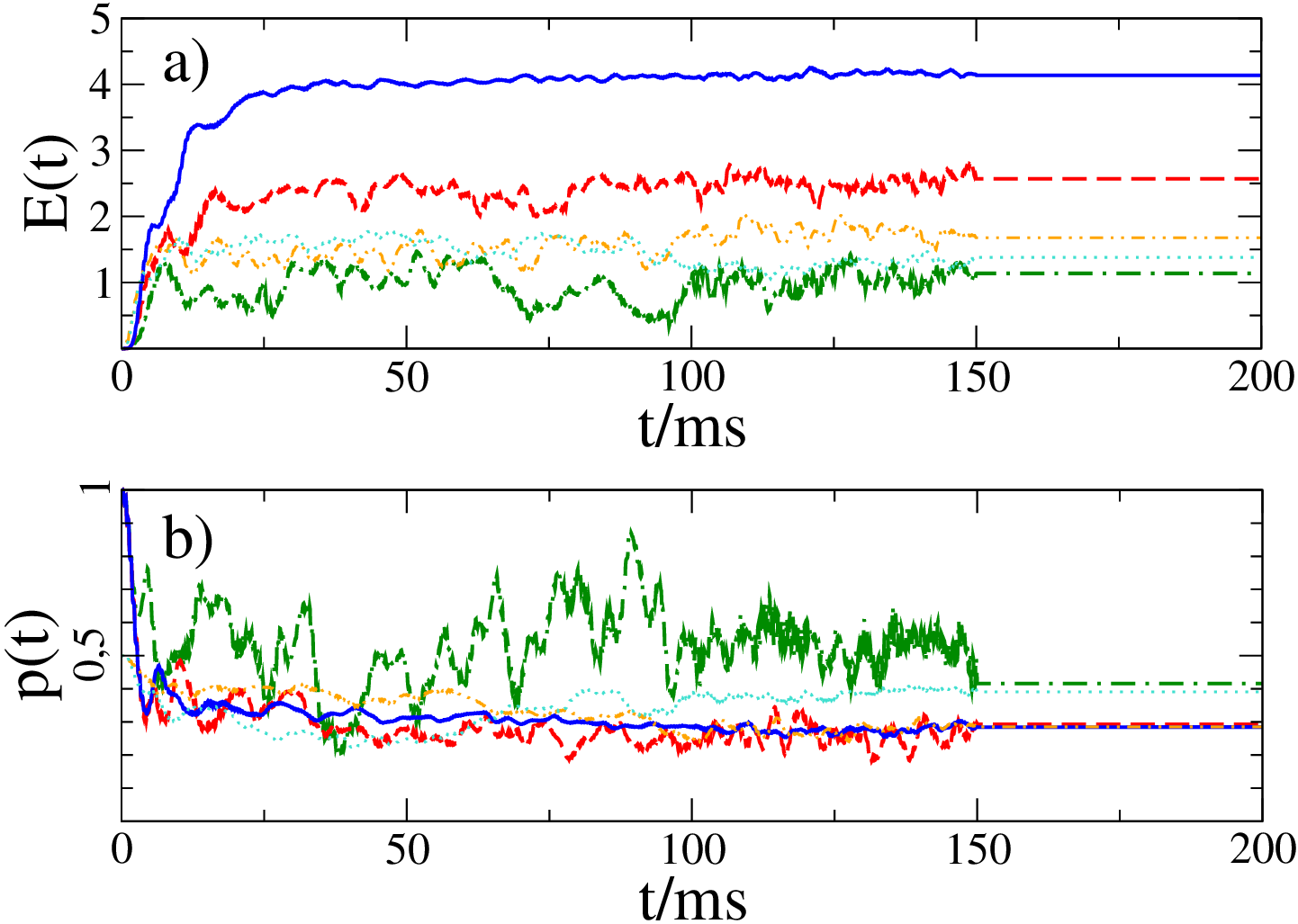}
\caption{\label{fig:eggsrob}(color online) Entropy of entanglement $E$ and probability of successful postselection for the driven Bose-Hubbard Hamiltonian \eqref{eq:V0} with \mbox{$L = 8 = N$} (solid blue), \mbox{$L = 6 = N$} (dashed red), \mbox{$L = 4 = N$} (dash-dotted green), \mbox{$L = 6$} and \mbox{$N = 5$} (thin dash-double-dotted orange), and \mbox{$L = 6$} and \mbox{$N = 7$} (thin dotted turquoise).
At \mbox{$t = 100 \,\mathrm{ms}$} the driving is stopped and at \mbox{$t = 150 \,\mathrm{ms}$} the lattice depth is increased in order to freeze the entanglement dynamics.}
\end{figure}

Fig.~\ref{fig:eggsrob} shows the dynamical enhancement of entanglement caused by the driving.
We start with the ground state of the static system.
After \mbox{$100 \,\mathrm{ms}$} the driving is switched off
and after \mbox{$t = 150 \,\mathrm{ms}$} the lattice depth is increased to \mbox{$V_{0} = 30 \,\mathrm{E_{R}}$} in order to freeze the entanglement dynamics completely.
Apparently, entanglement grows rather quickly once the driving is switched on.
After about $10$ to \mbox{$20 \,\mathrm{ms}$} the increase slows down a bit,
until entanglement saturates after \mbox{$t \approx 50 \,\mathrm{ms}$}.
From that time on, entanglement fluctuates around an average value due to the finite interactions in the system.
Fig.~\ref{fig:eggsrob} shows how these remaining fluctuations smooth out with growing system size.
Whereas the saturation time turns out to depend crucially on the lattice depth modulation $\mathrm{d}V$, the final values of entanglement seem not to.
Thus, a larger modulation of the lattice depth can accelerate the en\-tanglement generation;
however, even very weak driving can yield the same enhancement of entanglement as strong driving.

Once entanglement is saturated, it can be preserved with the help of a potential barrier that separates the two entangled atomic ensembles as depicted in Fig.~\ref{fig:model} and effectively switches off their mutual interaction.
The small fluctuations around the average entanglement that persist for \mbox{$t > 100 \,\mathrm{ms}$} can be stopped with an increase of the lattice depth, here at \mbox{$t = 150 \,\mathrm{ms}$}.

As it can be seen in Fig.~\ref{fig:eggsrob}a) the attainable entanglement grows with increasing system size.
This is as expected since larger systems can carry more entanglement.
A concern, however, is that with increasing number of atoms, also the number of possible distributions of atoms between the two subsystems is growing,
so that the probability to find an even distribution might decrease.
This decrease is apparent in Fig.~\ref{fig:eggsrob}b), where the probability of an even distribution drops from \mbox{$p = 0.33$} for $4$ atoms to \mbox{$p = 0.27$} for $5$, $7$ and $8$ atoms.
It also becomes ap\-parent, however, that this decrease occurs in small systems, and that these probabilities become largely independent of the particle number for large $N$.
Thus, one should expect to find evenly distributed particle numbers with substantial probability also in an experiment with significantly more bosons than a numerical simulation can handle.

\begin{figure}[h]
\centering
\includegraphics[width=0.45\textwidth]{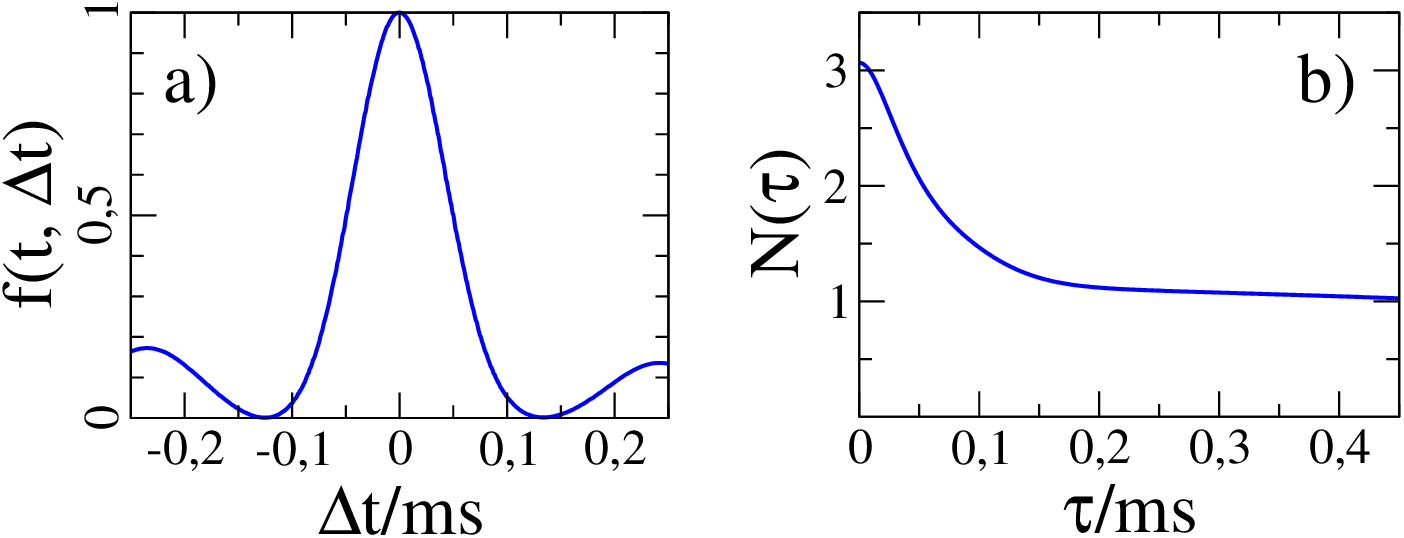}
\caption{\label{fig:rob23} a) Fidelity $f(t, \Delta t)$ at \mbox{$t = 100 \,\mathrm{ms}$} for state preparation with imperfect timing.
b) Negativity $N$ of the mixed state \eqref{eq:rho} resulting from the measurement of the driven Bose-Hubbard Hamiltonian with \mbox{$L = 6 = N$} at \mbox{$t_{0} = 100 \,\mathrm{ms}$}.}
\end{figure}

In an experiment, certainly also the timing will be crucial.
As Fig.~\ref{fig:eggsrob} shows, the system evolves rapidly, and
fluctuations in the durations of driving or ramping up the barrier that are comparable to system time scales will result in the generation of a mixed state which typically has reduced entanglement.
The relevant time scale can be obtained from the fidelity \mbox{$f(t, \Delta t) = |\langle \psi(t) | \psi(t+\Delta t) \rangle |^2$},
where $|\psi(t)\rangle$ is the postselected system state after driving of duration $t$.
The fidelity $f$ is depicted in Fig.~\ref{fig:rob23} for the exemplary case of \mbox{$N = 6$} particles in a \mbox{$L = 6$} site lattice.
The width of the central peak (full width at half maximum) that determines the minimal required experimental precision reads in this case \mbox{$\Delta t_{m} = 0.1 \,\mathrm{ms}$}.
In a similar fashion, we can also estimate the required precision for all other experimental parameters,
such as the potential $V$ (\mbox{$\Delta V_{m} = 0.08 \,\mathrm{E_{R}}$}),
the perpendicular confinement $V_{\perp}$ (\mbox{$\Delta V_{\perp, m} = 0.12 \,\mathrm{E_{R}}$}),
the amplitude of the driving $\mathrm{d}V$ (\mbox{$\Delta \mathrm{d}V_{m} = 0.016$}) and
the driving frequency $\omega$ (\mbox{$\Delta \omega_{m} = 14.5 \,\mathrm{Hz}$}).

To estimate the impact of fluctuations of these para\-meters on the attainable entanglement,
we have to consider the mixed state that is obtained with many repetitions of the experiment with the fluctuating parameter taking different values at each repetition.
To be specific, we focus on the
inaccuracy in the duration of driving,
and we assume that these durations are distributed according to a Gaussian centered around
\mbox{$t_{0} = 100 \,\mathrm{ms}$} with a standard deviation $\tau$.
This gives rise to the mixed state
\begin{eqnarray}\label{eq:rho}
 \rho(\tau) & = & \frac{1}{\tau \sqrt{2 \pi}} \int_{-\infty}^{\infty} \mathrm{d}t\
                   e^{-\frac{(t-t_{0})^{2}}{2 \tau^{2}}}
                   | \psi(t) \rangle \langle \psi(t) | \ .
\end{eqnarray}
The negativity of this mixed state is depicted in Fig.~\ref{fig:rob23}b)
as function of the inaccuracy $\tau$ of the duration of driving.
For \mbox{$\tau = 0 \,\mathrm{ms}$}, the situation reduces to the case of pure state entanglement as discussed above;
but for finite $\tau$ en\-tanglement is reduced significantly what is a very generic feature of mixed states.
Besides this expected behavior, there are two features that should be stressed:
\begin{itemize}
\item[]First, at \mbox{$\tau = 0 \,\mathrm{ms}$} the first derivative of $N(\tau)$ vanishes, so that entanglement turns out to be insensitive to small timing errors.
The second-order Taylor expansion reads \mbox{$N(\tau) \approx 3.064 - 900 ( \tau / \mathrm{ms} )^{2}$}.
That is, timing errors below $0.01 \,\mathrm{ms}$ imply a change in negativity of less than \mbox{$3 \,\mathrm{\%}$}.
\item[]Second, even in the presence of significantly larger timing errors there is still rather strong entanglement with \mbox{$N(\tau) \simeq 1$},
{\it i.e.}\ a value attained for a maximally entangled Bell pair!
\end{itemize}
In particular, this astonishing robustness against experimental fluctuations underpins that potential that driving offers as means to create entanglement as compared to engineered interactions.

As recent investigations on driven spin-systems suggest \cite{drivenions}, the feature of dynamical enhancement of entanglement is not particular for the Bose-Hubbard system,
but a rather generic feature, that is largely independent of detailed system properties.
An advantage of the present bosonic system as compared to many spin-systems is that particle numbers can easily be varied in an experiment,
what provides the means to study the generation of entanglement in the entire regime from rather small systems through the mesoscopic domain, up to the semi-classical regime.
In particular, observing the rise and decay of entanglement with increasing particle number will provide us with valuable insight in the emergence of classical behavior in large quantum systems.

M.L. thanks J. Eisert, M. B. Plenio and N. Schuch for very useful comments. S.W. acknowledges support by EMMI, DFG FOR760, and the Excellence Initiative through the Global Networks Mobility Measures, the Frontier Innovation Fonds and the HGSFP (DFG grant GSC 129/1).
F.M. acknowledges financial support by DFG (MI 1245-1).

\end{document}